\documentclass[12pt]{iopart}

\usepackage[]{graphicx}
\usepackage{amssymb}

\newcommand{\figwidth}{12cm}

\begin{document}
\title[Superconducting single-photon detector made of MoSi film]{Superconducting single-photon detector made of MoSi film}
\author{Yu P Korneeva$^1$, M Yu Mikhailov$^2$, Yu P Pershin$^5$, N N Manova$^1$, A V Divochiy$^3$, Yu B Vakhtomin$^3$, A A Korneev$^{1,4,6}$, K V Smirnov$^3$, A G Sivakov$^2$, A Yu Devizenko$^5$, and G N Goltsman$^{1,6}$}
\address{$^1$ Moscow State Pedagogical University, 1 Malaya Pirogovskaya str., 119991 Moscow, Russia}
\address{$^2$ B. Verkin Institute for Low Temperature Physics and Engineering of the National Academy of Sciences of Ukraine, 47 Lenina ave., 61103 Kharkiv, Ukraine}
\address{$^3$ CJSC ``Superconducting nanotechnology'' (Scontel), 5/22 Rossolimo str., 119991 Moscow, Russia}
\address{$^4$ Russian Quantum Center, 100 Novaya str., Skolkovo, Moscow Region}
\address{$^5$ National Technical University ``Kharkiv Polytechnic Institute'', 21 Frunze str., 61002 Kharkiv, Ukraine}
\address{$^6$ National Research University Higher School of Economics, 20 Myasnitskaya Ulitsa, Moscow 101000, Russia}
\ead{korneeva\_yuliya@mail.ru}

\begin{abstract}
We fabricated and characterised nanowire superconducting single-photon detectors (SSPDs) made of 4\,nm thick amorphous Mo$_x$Si$_{1-x}$ films. At 1.7\,K the best devices exhibit a detection efficiency up to 18\% at 1.2\,$\mu$m wavelength of unpolarised light
, a characteristic response time of about 6\,ns and timing jitter of 120\,ps. The detection efficiency was studied in wavelength range from 650\,nm to 2500\,nm. At wavelengths below 1200\,nm these detectors reach their maximum detection efficiency limited by photon absorption in the thin MoSi film.
\end{abstract}


\section{Introduction}
Recently the use of nanowire superconducting single-photon detectors (SSPDs)~\cite{APL01} was demonstrated in many applications ranging from quantum key distribution~\cite{WangOptLett12} to time-of-flight depth ranging~\cite{NatarajanSUST12}. While NbN was originally the first material used for SSPDs many other materials were tested during the past years such as Nb\,\cite{SemenovEurPhysJ05}, NbTiN\,\cite{DorenbosAPL08}, TaN\,\cite{EngelAPL2012}, MgB$_2$\,\cite{ShibataAPL10}, NbSi\,\cite{DorenbosAPL11}, W$_x$Si$_{1-x}$\,\cite{BaekAPL11}, and Mo$_{0.75}$Ge$_{0.25}$\,\cite{MarsiliMoGe14arXiv}. Among them the SSPDs made of amorphous W$_x$Si$_{1-x}$ (a-WSi) films turned out to be very promising. These SSPDs with also an improved optical coupling demonstrated at 120\,mK up to 93\% system detection efficiency at a wavelength near 1550\,nm\,\cite{MarsiliNatPhot13} and 75\% detection efficiency with 300 dark counts per second at 2.5\,K~\cite{FelixBussieresArXiv14}(supplementary material). Amorphous WSi is claimed to be attractive due to its critical temperature $T_c$ tunable over a wide range by the film stoichiometry, larger hot spot compared to NbN (for the same photon energy and film thickness), structural homogeneity and absence of grain boundaries. Due to high detection efficiency in the wavelength range of 1280 to 1650\,nm they are expected to be usable for the mid-infrared wavelengths~\cite{BaekAPL11}. Besides, amorphous films do not set strict requirements on the substrate crystal quality. This fact enables one to fabricate a cavity-integrated detector with the mirror placed on the substrate and front-side illumination of the detector (called ``Type I'' device in our previous work ~\cite{KorneevaIEEE13}).

Despite of the fact that by now a number of materials, other than NbN, have been studied, an optimally suitable material has not been found. For example, for some materials $T_c$ is too low. While the maximum reported $T_c$ of thick a-WSi films is 5~K~\cite{Kondo92}, the nanowire made of the 4.5 and 4~nm thick a-WSi film has a $T_c$ of 3~K~\cite{BaekAPL11} and 3.7~K~\cite{MarsiliNatPhot13}. For a-NbSi SSPDs a critical temperature near 2\,K was reported~\cite{DorenbosAPL11}. On another aspect, the single photon response of a-WSi SSPDs showed a saturation in internal detection efficiency at temperatures below 2\,K for 1.55\,$\mu$m radiation. However, the critical current of a-WSi SSPDs is only 2\,$\mu$A at 2\,K. In addition the best performing  devices were operated at a temperature as low as 120~mK (Fig.\,3 of~\cite{MarsiliNatPhot13}). In this paper we study SSPDs made of another refractory metal-silicon alloy: amorphous Mo$_x$Si$_{1-x}$ (a-MoSi). In comparison, the maximum $T_c$ of a-MoSi films is higher compared to the $T_c$ of a-WSi and a-NbSi. A $T_c$ of 7.5~K is reported for 25~nm thick a-Mo$_{0.75}$Si$_{0.25}$ film~\cite{SmithPRB94}, and of 7.3~K for a 50~nm thick a-Mo$_{0.80}$Si$_{0.20}$ film~\cite{KuboJAP88}. Using a-MoSi we intend to provide higher temperatures for efficient SSPD operation ($T\sim{1.7}$K) compared to a-WSi SSPDs, achievable by helium vapour pumping from a cryostat or a cryoinsert for storage Dewar as described in Korneev \textit{et al}~\cite{KorneevJSTQE07}. Our aim is to have excellent performance at temperatures above 1\,K. Therefore we have chosen to take a material with analogous properties as WSi but with a higher $T_c$. We expect in the temperature range of $T>$1\,K a higher critical current than in a-WSi, a lower response time and a lower timing jitter compared to a-WSi. Here we report our measurements of the detection efficiency, dark count rate, and timing characterisation of a-MoSi SSPD detectors made of films with Si content of 25\% and 20\% hereafter referred to as a-Mo$_{0.75}$Si$_{0.25}$ and a-Mo$_{0.80}$Si$_{0.20}$.

\section{Device fabrication and experimental set-up}
The 4\,nm thick a-MoSi films are deposited by dc magnetron sputtering from molybdenum and silicon targets on polished and thermally oxidised [100] Si wafers. SiO$_2$ layer is 250\,nm-thick. No intentional substrate heating is used and the substrate temperature does not exceed 100$^\circ$C during the sputter deposition. We evaluate the stoichiometry of the films from the calibrated deposition rates of Mo and Si. For constant settings of the deposition conditions the film thickness is controlled via the deposition time. The thickness of the films is checked by x-ray reflectivity measurements and their amorphous structure is confirmed by x-ray diffraction. The films are capped with a 3\,nm thick Si layer to prevent oxidation. Being partially oxidized in the natural way, this layer has a resistivity of about 2\,$\Omega\cdot$cm which is by 4 order of magnitude higher compared to a-MoSi resistivity. Thus we do not expect any influence of this caping layer on the electrical properties of a-MoSi. The patterning procedure is essentially the same as for the NbN detectors~\cite{FabricationOf}: the film is patterned into meanders covering 7\,$\mu$m\,$\times$\,7\,$\mu$m area by e-beam lithography in a positive-tone resist ZEP 520A7 and reactive ion etching in a SF$_6$ plasma. The meander strip width varied for different devices from 110\,nm to 130\,nm with the gap between the strips varying from 70\,nm to 90\,nm. After patterning of the meanders, the Cr/Cu  electrical contacts are fabricated by electron beam evaporation and patterned by photolithography method. Taking into account the large area of the contact pads (about 3\,mm$^2$) there are inevitably holes in the Si capping layer that provided a good contact between Cr/Cu and a-MoSi. Such  holes also might appear after ultrasonic bonding or by spring pins when the sample is mounted in the measurement dipstick.

The resistivity of a-MoSi is calculated from the sheet resistance $R_{s}$ and film thickness $d$. The sheet resistance is measured on a piece of wafer with unpatterned film by a four-point in-line probe technique using a commercial wafer probe: current $I$ is supplied by an outer pair of probes, voltage $U$ is measured across the two inner probes. If the probes are equidistant the sheet resistance can be calculated according to: $R_{s}=CU/I$,
where the correction factor $C$ accounts for the shape and size of the sample~\cite{Smits58}. The $R_{s}$ calculated from the sheet resistance of unpatterned film is in a reasonable agreement with $R_{s}$ calculated from patterned detector resistance, film thickness, nanowire width and length. Any observed descrepacies can be explained by inaccuracy in nanowire width measurement.

For detection efficiency ($DE$) measurement the detector is mounted in an dipstick placed inside a cryoinsert for a storage dewar enabling operation at 1.7\,K by evacuation of the helium vapour from the cryoinsert. The detector is illuminated with unpolarised light of 1.2\,$\mu$m wavelength from a light emitting diode (LED). The light is delivered to the detector by a bunch of multi-mode fibres which provide uniform illumination of the detector. Additional uniformity is achieved by placing the SSPD chip at a distance of about 1\,mm from the end of the fibre bunch. The SSPD is illuminated from the front side. The detection efficiency is defined as the ratio of photon counts to the number of incoming photons:
\begin{equation}
 DE=\frac{N_c - N_{drk}}{N_{ph}},
\end{equation}
where $N_c$ is the number of counts when the LED is on, $N_{drk}$ is the number of dark counts when the LED is off, $N_{ph}$ is the number of incoming photons. In our experiment the dark count rate with the LED  off is equal to the dark count rate with the fibre input  blocked by a black cap. 
To determine the number of photons incident on the detector we place a pinhole at the same position, where normally SSPD is located, and measure the optical power through the pinhole. Knowing the ratio of the areas of the pinhole and the a-MoSi meander we calculate the number of photons incident on the meander area. Since the radiation wavelength is several times larger than the gaps between the strips of the meander we do not take these gaps into account when we calculate the meander area, thus we take the whole area 7\,$\mu$m\,$\times$\,7\,$\mu$m. The results of this method are consistent with the results of detection efficiency measurements for SSPDs subsequently packaged with single-mode fibres.

In the procedure above we use a definite wavelength. To determine the wavelength dependence of the detection efficiency  we use a grating monochromator producing unpolarized light which is delivered to the SSPD by the same fibre bunch as we used for $DE$ measurement. The output power of the monochromator is calibrated with the Golay cell. All the rest of the setup is essentially the same as described above.

\section{Results and discussion}
\label{sec:results}
\subsection{Resistance vs temperature}
\begin{figure}
\includegraphics[width=\figwidth]{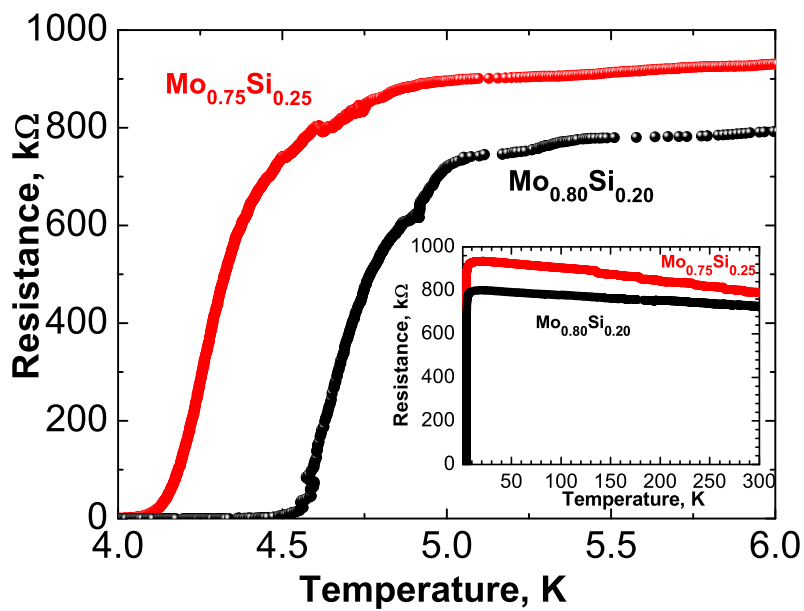}
\caption{\label{fig:RvsT} Comparison of the resistance vs temperature for SSPDs with 120-nm-wide strips made of Mo$_{0.75}$Si$_{0.25}$ (red) and Mo$_{0.80}$Si$_{0.20}$ (black). The inset shows the resistance vs temperature for the same devices in the temperature range up to 300\,K.}
\end{figure}

We measured the temperature dependence of the resistance for the meanders. We observe wide transition due to film disorder as shown in Fig.~\ref{fig:RvsT}. The devices and the unpatterned films have a small negative temperature coefficient of resistivity as shown in the inset of Fig.~\ref{fig:RvsT}. The ratio of the resistances at 300\,K and 20\,K temperatures $R_{300}/R_{20}$ is 0.85 and 0.90 for a-Mo$_{0.75}$Si$_{0.25}$ and a-Mo$_{0.80}$Si$_{0.20}$ devices respectively. The electrical resistivity of unpatterned films is about 220\,$\mu\Omega\cdot\mbox{cm}$ for a-Mo$_{0.75}$Si$_{0.25}$ and 185\,$\mu \Omega\cdot\mbox{cm}$ for a-Mo$_{0.80}$Si$_{0.20}$ at 10\,K which is in accordance with previously published data for amorphous a-MoSi~\cite{KuboJAP88} and a-WSi films~\cite{Kondo92}.

\subsection{Detection efficiency and dark count rate}
\begin{figure}
\includegraphics[width=\figwidth]{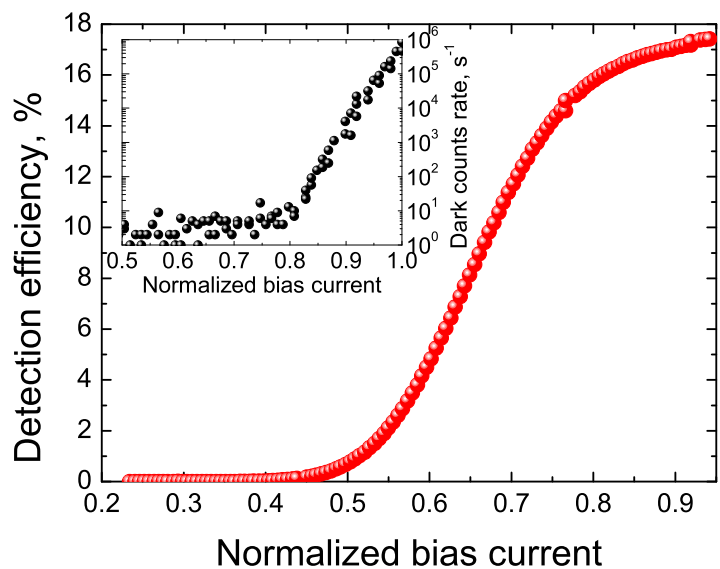}
\caption{\label{fig:qe} Detection efficiency for one of the best devices 
measured at 1.7\,K exhibits a beginning of a plateau at bias current close to $I_c$ ($I_c$=6.8\,$\mu$A). The inset shows dark count rate for the same device measured at the same temperature. The plateau in dark count at bias current below 0.8$I_c$ is caused by residual light and the electrical noise of the setup.}
\end{figure}

Figure~\ref{fig:qe} presents $DE$ for one of the best devices 
measured at 1.2\,$\mu$m wavelength of unpolarized light. This device showed critical current of 6.8\,$\mu$A at 1.7\,K. One can see that at bias current above 0.8$I_c$ detection efficiency exhibits somewhat like a beginning of a plateau, which is much better pronounced compared to NbN devices at the same wavelength and temperature, but yet it is not so clear as for WSi devices~\cite{BaekAPL11}.

The dark count rate 
is presented in the inset of Fig.~\ref{fig:qe}. The plateau at count rates below 10 counts per second and bias current below 0.8$I_c$ is caused by residual light and the electrical noise of the setup. The electrical noise is hard to filter due to poor signal-to-noise ratio.

\begin{figure}
\includegraphics[width=\figwidth]{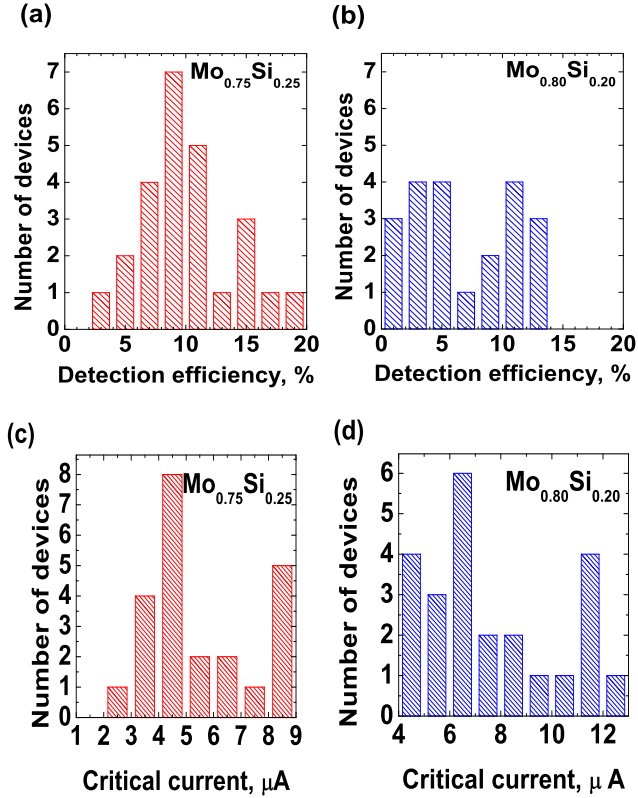}
\caption{\label{fig:starsky}Histograms of $DE$(a, b) and critical current (c,d)  measured for a-Mo$_{0.75}$Si$_{0.25}$ (a, c) and a-Mo$_{0.80}$Si$_{0.20}$ (b, d) devices at 1.7\,K.}
\end{figure}

%
%
Figure~\ref{fig:starsky} presents histograms of DE (a, b) and critical current (c,d)  measured for a-Mo$_{0.75}$Si$_{0.25}$ (a, c) and a-Mo$_{0.80}$Si$_{0.20}$ (b, d) devices at 1.7\,K. One can see that both materials are quite similar. Devices made from  a-Mo$_{0.80}$Si$_{0.20}$ film exhibit higher critical currents which may be caused by higher $T_c$.  Although one can see that a-Mo$_{0.75}$Si$_{0.25}$ devices exhibit better fabrication yield and higher detection efficiency we do not have enough statistical data to claim that it is the property of material itself and not the imperfection during the fabrication process.

\begin{figure}
\includegraphics[width=\figwidth]{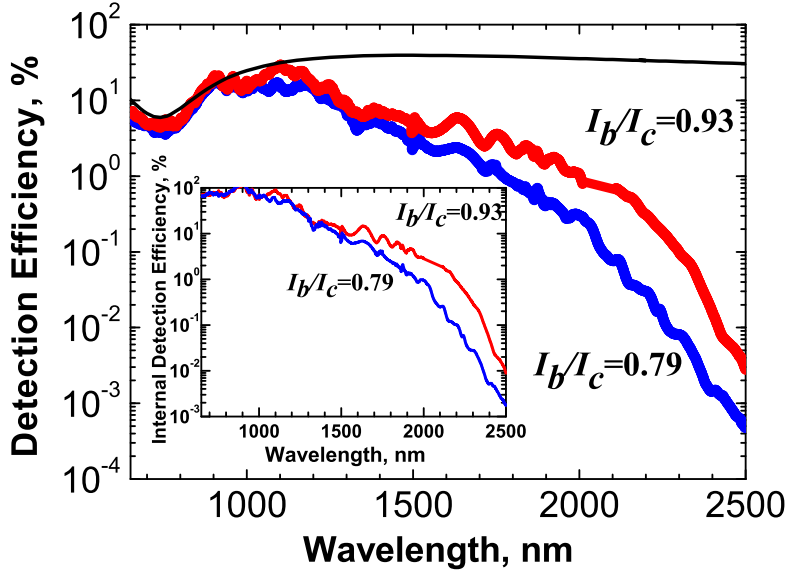}
\caption{\label{fig:spectrum}Detection efficiency vs wavelength measured at bias currents 0.93$I_c$ (red curve) and 0.79$I_c$ (blue curve). Top black curve shows simulated absorption of multilayer stack of 3-nm capping Si, a-MoSi detector, SiO$_2$ and Si. The inset shows internal detection efficiencies of a-MoSi SSPD at bias currents 0.93$I_c$ and 0.79$I_c$.}
\end{figure}

Figure~\ref{fig:spectrum} presents detection efficiency vs wavelength for a-Mo$_{0.75}$Si$_{0.25}$ SSPD in range 650\,nm to 2500\,nm at bias currents 0.93$I_c$ (red points) and 0.79$I_c$ (blue points). The top black curve is the simulated absorption of multilayer stack consisting of 3-nm-thick capping Si, a-MoSi detector, 250-nm-thick SiO$_2$, and Si. The simulation was performed with 1D array transfer matrix method~\cite{YehOpticalWavesInLayeredMedia}. Optical refractive indices of the dielectric materials (Si and SiO$_2$) were taken from~\cite{refractiveindex}, whereas reflectance of patterned a-MoSi film was calculated from its sheet resistance at 20\,K. Our simulation showed that 3\,nm capping does not significantly affect the absorption but shifts the wavelengths of maximum and minimum absorption by several nanometers. One can note that at wavelengths shorter than 1000\,nm $DE$ is very close to the simulated absorption indicating almost 100\% internal $DE$, i.e. every absorbed photon produces a count. This is clearly indicated in the inset of Fig.~\ref{fig:spectrum} where we plotted internal $DE$ which is the ratio of measured $DE$ to simulated absorption. A plateau up to 1200\,nm wavelength is clearly visible in the internal $DE$ at both bias currents.

\subsection{Response time and timing jitter}
\begin{figure}
\includegraphics[width=\figwidth]{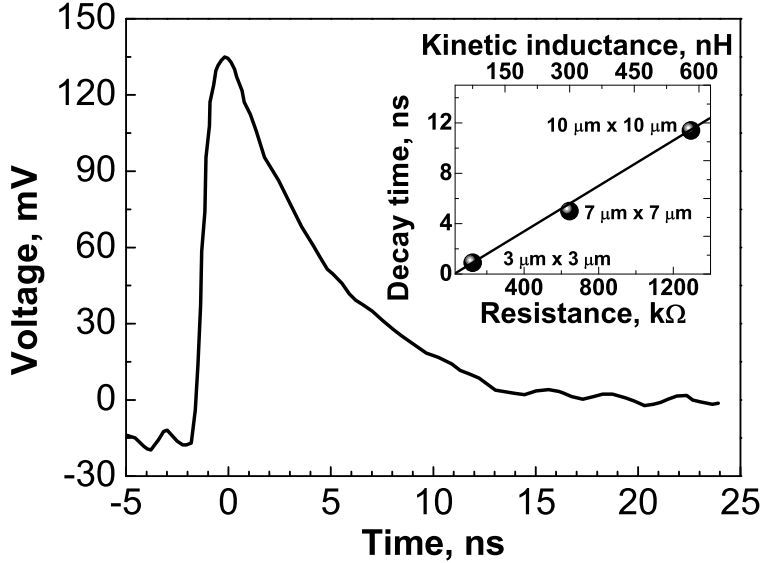}
\caption{\label{fig:waveform} Averaged waveform transient for 7\,$\mu$m\,$\times$\,7\,$\mu$m SSPD. The inset shows the dependence of the response fall time on the detector room temperature resistance. The linear dependence supports the idea that the response time is limited by the kinetic inductance.}
\end{figure}

Figure~\ref{fig:waveform} presents a typical averaged waveform transient for 7\,$\mu$m\,$\times$\,7\,$\mu$m area a-MoSi 
detector after about 46\,dB amplification by 2 amplifiers Mini-Circuits ZFL-1000LN+ with 0.1--1000 MHz band and noise figure of about 6\,dB. We evaluate the influence of the  kinetic inductance on the response fall time by fabricating meanders of two more sizes: 3\,$\mu$m\,$\times$\,3\,$\mu$m, and 10\,$\mu$m\,$\times$\,10\,$\mu$m.  As it was shown in~\cite{KermanNbNReset} the response fall time $\tau_f$ of SSPD is limited by the kinetic inductance of the wire $L_k$: $\tau_f=L_k/R$, where $R$ is the impedance of the coaxial transmission line which is in most cases (as well as in our case) 50~$\Omega$. Here we define the response time as the time required for the voltage to decay by a factor of $e$ (the base of the natural logarithm). We did not measure $L_k$ directly. Instead we followed the approach proposed in~\cite{KermanNbNReset} and used the fact that $L_k$ has the same dependence on strip length $l$, width $w$ and film thickness $d$ (~\cite{TinkhamIntroToSupercond2ndEdition}) as room temperature resistance $R_{300}$:
\begin{equation}
L_k = \mu_0 \lambda^2 \int_0^l \frac{dx}{w(x) d(x)}, \mbox{~and~} R_{300}=\rho \int_0^l \frac{dx}{w(x) d(x)},
\end{equation}
where $\lambda$ is magnetic field penetration depth, $\rho$ is the resistivity of the film material at room temperature. This approach gives a better accuracy because even if there is a variation of $l$, $w$, and $d$, $L_k$ remains proportional to $R_{300}$.
%
%

The inset in Fig.~\ref{fig:waveform} shows the response fall time vs room temperature resistance for the meanders of different lengths. One can see that the response time is close to a linear dependence suggesting that even for 3\,$\mu$m\,$\times$\,3\,$\mu$m devices the recovery time is still limited by the kinetic inductance. Taking into account the geometry of our devices we calculated a kinetic inductance of a square unit of our device $L_{\Box}$=0.15\,nH. This is rather close to NbN meanders ($L_{\Box}$=0.12\,nH for SSPD reported in~\cite{GoltsmanJMO09}) and approximately by a factor of 2 smaller than $L_{\Box}$ of WSi reported in~\cite{MarsiliNatPhot13} (taking 120\,ns decay time between 90\%--10\% levels, 15\,$\mu$m$\times$\,15\,$\mu$m detector size, 120\,nm strip width and 200\,nm pitch we obtained $L_{\Box}$=0.3\,nH for WSi reported in~\cite{MarsiliNatPhot13}).

\begin{figure}
\includegraphics[width=\figwidth]{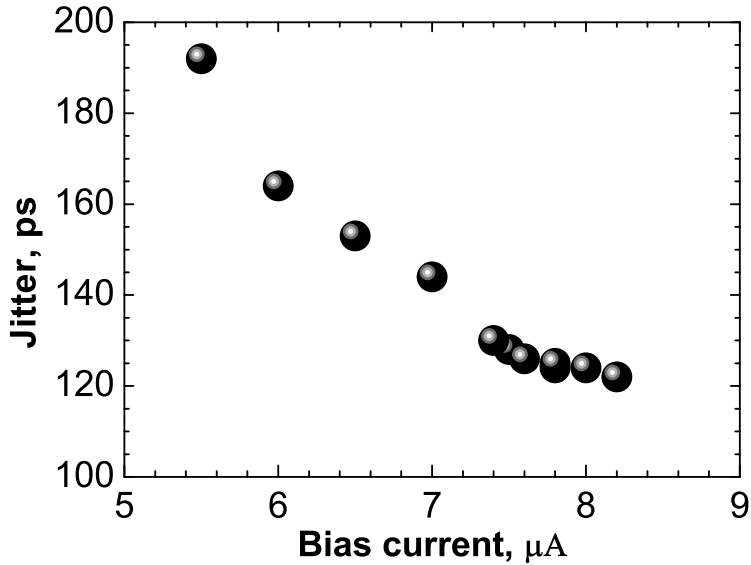}
\caption{\label{fig:jitter} Timing jitter of a-MoSi SSPD measured at different bias currents. Such a dependence of the jitter on bias current together with low response voltage of the detector suggest that in our case jitter is limited by the noise of the read-out electronics similar to~\cite{LixingYouAIPAdvances13}.}
\end{figure}

We determine the timing jitter of the a-MoSi SSPDs from the width of the coincidence histogram taken at the half of its maximum. The jitter is measured for different bias currents ranging from 5.5~$\mu$A to $I_c$ which is about 8.3~$\mu$A for this device. The results are presented in Fig.~\ref{fig:jitter}. The timing jitter is decreasing from 190\,ps at 5.5\,$\mu$A bias current to about 120\,ps near $I_c$. Such a dependence on bias current is very similar to the one reported for NbN by You et\,al~\cite{LixingYouAIPAdvances13} and for WSi in supplementary material of~\cite{MarsiliNatPhot13}. The agreement of our data and their data suggests that in our case timing jitter is also limited by the noise of the read-out. Indeed, when $I_c<10$~$\mu$A the response voltage of the detector is just several times higher than the RMS of the amplifier noise voltage. Although these values of the jitter are not the properties of the material nevertheless they give an idea of what one can obtain in a practical MoSi single-photon receiver operated with room-temperature amplifiers.

\section{Conclusion}
We fabricated and tested amorphous a-Mo$_x$Si$_{1-x}$ SSPDs. These devices exhibited critical temperatures of 4.3--4.9\,K allowing their successful operation at 1.7\,K temperature. The best devices have maximum detection efficiency of 18\% at 1.2\,\,$\mu$m wavelength for unpolarized light, and 16\% at  10 dark counts per second. We demonstrate a kinetic inductance limited 1/$e$ voltage decay time for a 7\,$\mu$m\,$\times$\,7\,$\mu$m detector of about 6\,ns with a timing jitter of 120\,ps limited by the RMS of the amplifier noise.

Although the best detection efficiency we achieve so far is about 18\% we believe that this material has high potential for future SSPD development, because it allows the integration with optical cavities enabling front-side illumination. 

\ack
M.Yu.M. would like to thank~A.~S.~Garbuz, whereas Yu.P.K. would like to thank ~T.~M.~Klapwijk for valuable discussions on device physics and a critical reading of the manuscript. This work was supported in part by Russian state Contract 14.B25.31.0007 from the Ministry of Education and Science of the Russian Federation, by grant 26/13-H of the targeted program ``Fundamental problems of nanostructure systems, nanomaterials and nanotechnologies'' of NAS of Ukraine, and Russian Quantum Center.

\section*{References}

\bibliography{allrefs}
\bibliographystyle{unsrt}

\end{document}